\documentclass[12pt]{article}
\usepackage[pdftex]{color}
\usepackage{amsmath,amsthm,amssymb}
\usepackage{wasysym}
\usepackage{graphicx}
\usepackage{color}%
\usepackage[onehalfspacing]{setspace}
\usepackage{lipsum}
\usepackage[lmargin=2.5cm,rmargin=2.5cm,tmargin=3cm,bmargin=3cm]{geometry}
\usepackage{multirow}
\usepackage{hyperref}
\hypersetup{hidelinks}
\usepackage{cite}
\usepackage{cleveref}
\usepackage{accents}
\usepackage[ruled,vlined,linesnumbered]{algorithm2e}
\usepackage[numbers,sort&compress]{natbib}
\usepackage{scalerel}
\usepackage{stackengine}
\usepackage{booktabs}
\usepackage[ruled,vlined,linesnumbered]{algorithm2e}
\stackMath
\definecolor{color}{rgb}{0.8500, 0.3250, 0.0980} 

\usepackage{nomencl}
\makenomenclature
\usepackage{etoolbox}
\renewcommand\nomgroup[1]{%
  \item[\bfseries
  \ifstrequal{#1}{A}{Variables}{%
  \ifstrequal{#1}{B}{Superscripts}{%
  \ifstrequal{#1}{C}{Subscripts}{%
  \ifstrequal{#1}{D}{Abbreviations}{
  \ifstrequal{#1}{E}{Special functions}{
  \ifstrequal{#1}{F}{Operators}{}}}}}}%
]}

\begin{document}
\begin{center}
{\rm \bf \Large{Electrified EHL line contact with dielectric breakdown of lubricant - a numerical model}}
\end{center}

\begin{center}
{\bf Yang Xu}$^{\text{ab}}$, {\bf Nick Morris}$^{\text{c}}$\footnote{Corresponding author: n.j.morris@lboro.ac.uk}, {\bf Yue Wu}$^{\text{d}}$\footnote{Corresponding author: wuy@sari.ac.cn}
\end{center}
\begin{flushleft}
{
$^{\text{a}}$School of Mechanical Engineering, Hefei University of Technology, Hefei, 230009, China \\
$^{\text{b}}$Anhui Province Key Laboratory of Digital Design and Manufacturing, Hefei, 230009, China \\
$^{\text{c}}$Wolfson School of Mechanical, Electrical and Manufacturing Engineering, Loughborough University, UK \\
$^{\text{d}}$Shanghai Advanced Research Institute, Chinese Academy of Sciences, Shanghai 201210, China
}
\end{flushleft}
\begin{abstract}
With the rapid growth of the electric vehicles with drive systems with higher voltages, power outputs, frequencies, and speeds, mitigating electrically induced bearing damage (EIBD) in electric motors has become critical. In this study, a novel numerical model characterizing discharge-induced current density and voltage drop at the elastohydrodynamic lubrication line contact interface is presented. The current density and voltage drop constitute a linear complimentarily problem, which is efficiently solved using the conjugate gradient method. This paper sheds light on electrical characteristics at the inaccessible lubrication interface during discharge, highlighting the significance of roughness radius of curvature on current density. This numerical model lays the groundwork for future research on mitigating or even permanently solving EIBD problems in electric motor bearings. 
\end{abstract}


\section{Introduction}
Given the growing concerns about climate change, the global shift toward electricity as the primary energy source for light-duty transportation, including electric vehicles (EVs), wheel loaders, buses, is accelerating. The electric motor, which serves as the core power unit for electrified transportation, is continually constrained by electrically induced bearing damage (EIBD), which affects the stability and reliability of the driven end and non-drive end bearings \cite{he2020electrical, chen2020performance}. In the EV industry at present, electric motor maximum speeds have already exceeded 20,000 RPM, and some are approaching 30,000 RPM. At such a high speed, the lubricant film between the rolling element and the inner/outer raceway becomes extremely thin due to a relatively high temperature, thermo-viscous, and shear-thinning behaviour. The lubricant film is subjected to high-frequency alternating shaft voltage or current, and the dielectric breakdown of the lubricant film is likely to occur intermittently near the minimum film thickness \cite{morris2022electrical,guo2023study}. When the dielectric strength of lubricant is exceeded, a plasma channel is formed between the surfaces. Through heat flux and charged particles (electrons and ions), driven by the electric field, bombarding the contiguous surfaces, pits and craters of various sizes and depths \cite{schneider2022electrical} similar to that created by electrical discharge machining (EDM) are formed \cite{Dhanik05}. Eventually, it causes EIBD on the surfaces of rolling elements and raceways, resulting in pitting, frosting, fluting, and other forms of surface damage \cite{he2020electrical,chen2020performance, janik2024exploring, bond2024influence,tajedini2024influence,guo2023study,yousuf2025influence,Notay25}.

EIBD is not a new problem; it has been widely prevalent in various fields, such as turbomachinery, wind power generation, and rail transit \cite{vance1987electric}, long before the emergence of the modern EV industry. Recently, there has been a renewed focus on EIBD from industry and academia, given the rapid growth of the EV market, particularly in China, and the move toward electric powertrains with higher voltage architectures ($400$V to $800$V), power outputs, pulse width modulation switching frequencies, and rotor speeds. Several solutions have been proposed to address the EIBD, including slip rings \cite{he2020electrical}, conductive grease \cite{suzumura2016prevention,bond2024influence,janik2024exploring,janik2024rheological,zhou2024electric}, and hybrid and full ceramic bearings \cite{oliver2015ceramic}. Some EV manufacturers have incorporated these solutions into their products. However, the electrical brushes in the slip rings are in frictional contact with the rotor and as a result of wear require regular servicing \cite{xiao2023sliding,argibay2010asymmetric}. The conductivity of the conductive grease gradually decreases with time. Therefore, considering both cost and system reliability, other EV manufacturers live with EIBD without taking any prevention. As a result the prevalence of EIBD could soon be an important factor influencing EVs' reliability and driving comfort.

To delay or even prevent EIBD in electric motor bearings, a thorough understanding of the electrified lubricated interface is required through testing and modeling. Experimental studies of EIBD have advanced in the last decade with electrical discharge events investigated at vehicle level (system level) \cite{LIU2025205863}, an electric motor (sub-system level) \cite{hwang2025bearing}, a rolling element bearing (component level) \cite{bleger2024automotive,sanchez2024rolling,graf2020surface,gonda2019influence}, or a lubricant interface \cite{guo2023study,sunahara2008development,sunahara2011preliminary,bond2024influence,lee2024electrification,janik2024exploring,janik2024rheological,li2024study,luo2006gas,yousuf2025influence,farfan2025influence, farfan2025electrified}. Since discharge events can occur at various coupled, inaccessible electrified interfaces, it is convenient to study discharge events at a single electrified interface of a ball-on-disk kinematic pair. Sunahara \cite{sunahara2008development,sunahara2011preliminary} measured the discharge locations and the corresponding lubricant film thickness using high-speed interferometric imaging techniques. The author found that discharge events primarily occur at the location of the minimum film thickness, specifically at the two side lobes of the horseshoe-shaped film thickness. A characteristic wear pattern, consisting of two parallel wear bands, was observed at the electrified ball-on-disk interface, rather than the well-known washboard pattern typically found in bearings. Sunahara's important observations have been confirmed independently by Guo et al. \cite{guo2023study,guo2024investigation}, who found two parallel wear bands composed of densely distributed EDM-induced pits. The distance between the center lines of two parallel bands is identical to the distance between two minimum film thickness locations. Guo et al. \cite{guo2023study,guo2024investigation} found that the EIBD severity first increases and then decreases with increasing minimum film thickness, with the maximum severity occurring at $251$ nm. Li et al. \cite{li2024study} conducted a similar ball-on-disk lubrication test, which resulted in an unexpected observation that the EIBD becomes worse at the hydrodynamic lubrication regime with a higher film thickness. They hypothesized that it may be linked to the formation of micro-bubbles in the electric field \cite{luo2006gas}. A recent study by Yousuf et al. \cite{yousuf2025influence}  pointed out that the surface electrical erosion is due to a complex competition between the electric field, tribofilm formation, polarity, and evolution of erosion. 
 
Currently, no models exist for simulating discharge events over the lubricated interface. As a compromise, the lubrication interface with discharge is commonly simplified as a resistor and a capacitor in parallel \cite{magdun2009calculation,gemeinder2014calculation,schneider2021empirical,turnbull2023electrotribodynamics,bleger2024automotive}. The bulk resistivity, solid-solid contact, and discharge-induced conductive path can all contribute to the magnitude of the resistor. The bulk resistivity of the dielectric lubricant or grease is so large that the interface can be considered an open circuit if there are no solid- solid contact and discharge events. The capacitance is proportional to the integral of the reciprocal of the lubricant film thickness over the entire non-conductive lubricated interface \cite{ten1970influence,schneider2021empirical}. Tracing the Stribeck curve from the hydrodynamic regime to the boundary lubrication regime, the discharge of the lubricant film thickness transits from a capacitive type to a resistive type \cite{chen2020tribological,graf2020surface}. However, most of the circuit models do not consider the influence of discharge events on the resistor, capacitor, and impedance values. Bledger et al. \cite{bleger2024automotive} assumed that the resistance of the discharging film is $10$ Ohm, and it is independent of the applied voltage/current, minimum film thickness, and dielectric strength \cite{chen2020tribological}.  
The pioneering work of Chatterton et al. \cite{chatterton2016electrical} numerically modeled the evolution of the topography of the electrically eroded surface of one tilting pad in a thrust bearing. The pitting location and corrosion depth are determined based on a phenomenological model regarding temperature, pressure, and film thickness. The coupling between the EIBD and lubrication was successfully implemented by Chatterron et al. in a commercial finite element software package. Jackson et al. \cite{jackson2024statistical} have recently developed an electrified mixed lubrication model where the probability of a discharge zone can be estimated for varying operating conditions and properties, assuming that the lubricant has a constant dielectric strength. Xu et al. \cite{xu2024} proposed a numerical framework that enables the numerical solution of the discharge-induced current density and voltage drop across the dry electrical contact interface. This work opens up a new avenue for quantitatively studying the discharging events within the lubricant without resorting to the complex particle collision model \cite{go2014microscale}. Two analytical solutions are derived for two cases where two parabolic electrodes are in contact and completely separated.

This study focuses on understanding EIBD in full film EHL lubrication combining the electrified EHL line contact model developed by Morris et al. \cite{morris2022electrical} with the discharge model proposed by Xu et al. \cite{xu2024}. Section 2 introduces all the essential equations for steady-state, isothermal EHL line contact. The formulation of the electrified EHL line contact with discharge, along with the corresponding numerical model, is presented in Section 3. Section 4 thoroughly investigates the impact of discharge on the electrical properties of the EHL line contact. The effect of roughness on the electrical properties of discharge is discussed in Section 5.

\section{EHL line contact}

\subsection{Problem statement}
\begin{figure}[h!]
  \centering
  \includegraphics[width=9cm]{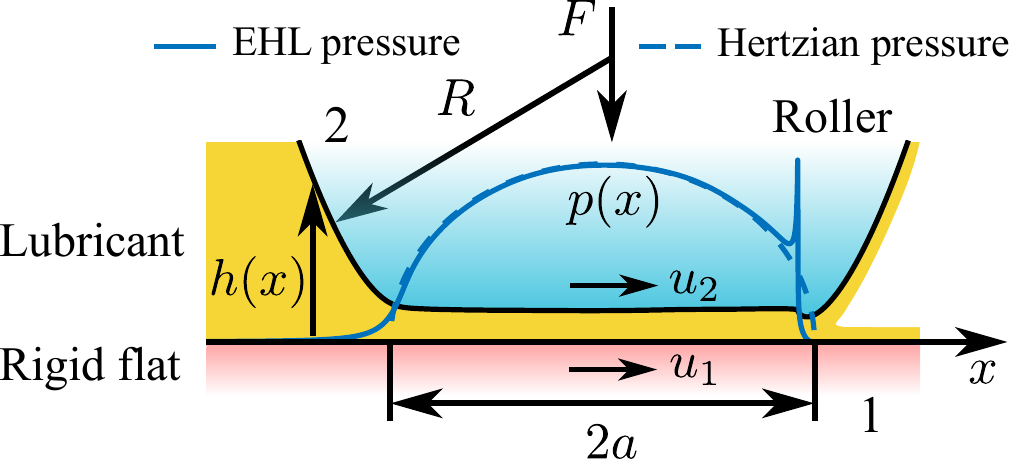}
  \caption{Schematic of a typical EHL line contact}\label{fig:Fig_1}
\end{figure}

Consider an elastohydrodynamic lubrication (EHL) line contact between the inner race of radius $R_1$ and a roller of radius $R_2$, see Fig. \ref{fig:Fig_1}. Both surfaces are elastic with Young's modulus $E_i$ and Poisson's ratio $\nu_i$, $i = 1, 2$. The line contact problem can be simplified to a rigid flat in lubricated contact with an elastic roller with an equivalent radius $R$ and a reduced elastic modulus $E'$,
\begin{equation}
1/R = 1/R_1 + 1/R_2, ~~~ 2/E' = (1 - \nu_1^2)/E_1 + (1 - \nu_2^2)/E_2.
\end{equation}
Both surfaces translate horizontally with the velocities of $u_i$, $i = 1, 2$; The pressure and lubricant film thickness are denoted by $p(x)$ and $h(x)$, respectively. The lubricated contact is subjected to the normal force per unit axial length $F$. At high load conditions, the EHL line contact can be approximated by the Hertzian contact theory\cite{johnson1987contact},
\begin{equation}
p_{\text{h}} = \frac{2 F}{\pi a}, ~~~ a^2 = \frac{8 F R}{\pi E'},
\end{equation}
where $p_{\text{h}}$ is the maximum of the Hertzian pressure distribution, and $a$ is the semi-width of the contact area.

\subsection{Continuous equations}
The Reynolds equation of steady-state line contact is 
\begin{equation}
\frac{\partial}{\partial x} \left( \frac{\rho h^3}{12 \eta} \frac{\partial p}{\partial x} \right) - u_{\text{m}} \frac{\partial (\rho h)}{\partial x} = 0,
\end{equation}
where $\rho$ is the lubricant density, $u_{\text{m}} = (u_1 + u_2)/2$ is the speed of entrainment, and $p$ is the fluid film pressure. The film thickness equation is
\begin{equation}\label{eq:dimensional_film}
h(x) = h_0 + \frac{x^2}{2 R} - \frac{2}{\pi E'} \int_{-\infty}^{\infty} p(\xi) \ln |x - \xi|^2 \text{d}\xi,
\end{equation}
where $h_0$ is the rigid body approach. The isothermal viscosity to pressure relation of the lubricant is characterized by Roelands' equation,
\begin{equation}
\eta(p) = \eta_0 \exp\left[ (\ln(\eta_0) + 9.67) \times (-1 + (1 + p/p_0)^z)\right],
\end{equation}
where $p_0 = 1.96 \times 10^8$ in Pascals (Pa), $\eta_0$ is the viscosity at the ambient pressure, and $z$ is a constant. The density-to-pressure relation of the lubricant is characterized by the Dowson and Higginson equation,
\begin{equation}
\rho(p) = \rho_0 \left( 5.9 \times 10^8 + 1.34 p \right)/\left(5.9 \times 10^8 + p\right),
\end{equation}
where $\rho_0$ is the density of the lubricant at the ambient pressure. The lubricant pressure distribution should balance the external load, i.e., 

\begin{equation}
\int_{-\infty}^{\infty} p(x) \text{d}x = F.
\end{equation}
Using the following dimensionless group: 

\begin{align}\label{eq:dimensionless_group1}
x^* &= x/a, ~~~ h^* = h R/a^2, ~~~ p^* = p/p_{\text{h}}, ~~~ h_0^* = h_0 R/a^2, \notag \\
\eta^* &= \eta/\eta_0, ~~~ \rho^* = \rho/\rho_0, ~~~ W = F/E' R, ~~~ U = \eta_0 u_{\text{m}}/(E' R), 
\end{align}
the dimensionless Reynolds equation is 
\begin{equation}\label{eq:Reynolds_dimensionless}
\frac{\partial}{\partial x^*} \left( \xi \frac{\partial p^*}{\partial x^*}\right) - \frac{\partial (\rho^* h^*)}{\partial x^*}  = 0,
\end{equation}
where $\xi = \rho^* \left(h^*\right)^3/(\eta^* \lambda)$ and $\lambda = 12 u_{\text{m}} \eta_0 R^2/(a^3 p_{\text{h}})$. Similarly, the dimensionless film thickness and other dimensionless equations are tabulated below: 
\begin{equation}\label{eq:Dimensionless_film}
h^*(x^*) = h_0^* + \left(x^*\right)^2/2 - \frac{1}{\pi} \int_{-\infty}^{\infty} p^*(s) \ln|x^* - s| \text{d}s,
\end{equation}
\begin{equation}
\eta^*(p^*) = \exp\left[ (\ln(\eta_0) + 9.67) \times (-1 + (1 + p_{\text{h}} p^*/p_0)^z)\right],
\end{equation}
\begin{equation}
\rho^*(p^*) = \left(5.9 \times 10^8 + 1.34 p_{\text{h}} p^*\right)/\left(5.9 \times 10^8 + p_{\text{h}} p^*\right),
\end{equation}
\begin{equation}
\int_{-\infty}^{\infty} p^*(x^*) \text{d} x^* = \frac{\pi}{2}.
\end{equation}

\subsection{Discretized equations}\label{subsec:Discrete_EHL}
The computational domain is defined as, $\Omega = \{x^* | x_{\text{a}}^* \leq x^* \leq x_{\text{b}}^*\}$, which is uniformly discretized using $n$ nodes with nodal coordinate $x_i^* = x_{\text{a}}^* + (i - 1) \Delta_x^*$, $i = 1, \cdots, n$, where $\Delta_x^* = (x_{\text{b}}^* - x_{\text{a}}^*)/(n-1)$. Using central and backward differences, respectively, for the first and second terms in the dimensionless Reynolds equation, its discretized form at non-boundary nodes ($i \in [2, n-1]$) is as follows: 
\begin{equation}
\frac{1}{\left(\Delta_x^*\right)^2} \left[ \xi_{i+1/2} p_{i+1}^* - (\xi_{i + 1/2} + \xi_{i - 1/2})p_i^* + \xi_{i - 1/2} p_{i - 1}^* \right] - \frac{1}{\Delta_x^*} \left(\rho_i^* h_i^* - \rho_{i-1}^* h_{i-1}^* \right) = 0,
\end{equation}
where $\xi_{i \pm 1/2} = (\xi_i + \xi_{i \pm 1})/2$. The discretized film thickness is
\begin{equation}
h_i^* = h_0^* + \left(x_i^*\right)^2/2 - \frac{1}{\pi} \sum_{j = 1}^n K_{ij}^* p_j^*,
\end{equation}
where the dimensionless influence coefficient $K_{ij}^*$ has the closed-form as follows \cite{venner1991}:
\begin{equation}\label{eq:IC}
K_{ij}^* = (x_i^* - x_j^* + \Delta_x^*/2)( \ln|x_i^* - x_j^* + \Delta_x^*/2| - 1) - (x_i^* - x_j^* - \Delta_x^*/2)( \ln|x_i^* - x_j^* - \Delta_x^*/2| - 1).
\end{equation}
The discretized load equilibrium is
\begin{equation}
\frac{\pi}{2} = \frac{\Delta_x^*}{2} \sum_{i = 1}^{n-1} (p_i^* + p_{i+1}^*).
\end{equation}
The EHL line contact problem described above has been extensively studied \cite{venner1991,venner2000multi} and is solved by the multigrid method in the present study. It is implemented in MATLAB, following the details provided by Venner (Chapters 3 and 4 in Ref. \cite{venner1991}). 
 
\section{Electrified EHL line contact with discharge}
Xu et al. \cite{xu2024} developed a numerical model for electrical contact between electrodes with discharge events. This model can be easily adapted to characterize the electrical properties of the electrified lubrication interface perturbed by discharge events. In the remainder of this section, a model of the electrified EHL line contact subjected to a constant applied current or voltage using Ampère’s law is presented. 

\begin{figure}[h!]
  \centering
  \includegraphics[width=16cm]{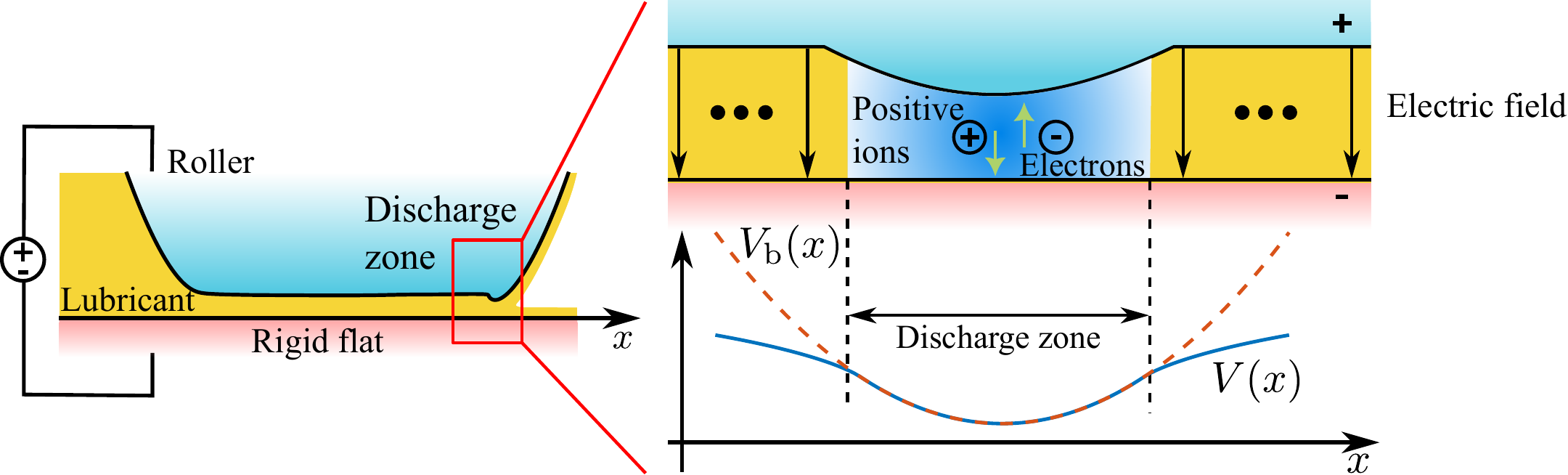}
  \caption{Schematic of electrified EHL line contact interface and discharge zone}\label{fig:Fig_2}
\end{figure}

\subsection{Governing equations}\label{subsec:Discharge_model}
The resistivities of the roller and rigid flat are $\rho_1$ and $\rho_2$. Let the potential drop and the current density at the interface be $V(x)$ and $J(x)$. The current per unit axial length of the roller, $I$, passes across the thin lubricant film. If $V(x)=V_{\text{b}}$ (the dielectric breakdown voltage of the lubricant), a conductive path with $J(x) > 0$ is formed perpendicular to the lubrication interface. If $V(x) < V_{\text{b}}(x)$, the lubricant film at that specific location remains non-conductive and $J(x) = 0$. Therefore, the interfacial discharge is described using the Karush-Khun-Tucker condition in terms of $\widetilde{V}(x) = V_{\text{b}}(x) - V(x)$ and $J(x)$:
 
\begin{alignat}{3}
&\widetilde{V}(x) = 0, ~~~ &&J(x) > 0, ~~~ && x \in \Omega_{\text{con}}, \label{eq:electric_cont} \\
&\widetilde{V}(x) > 0, ~~~ &&J(x) = 0, ~~~ && x \in \Omega_{\text{ncon}}, \label{eq:electric_ncont} 
\end{alignat}
where $\Omega_{\text{con}}$ and $\Omega_{\text{ncon}}$ are the conductive and non-conductive regions, respectively. 

Assuming the dielectric strength of the lubricant film is a constant, the breakdown voltage based on the parallel-plate capacitor theory can be estimated as
\begin{equation}\label{eq:Vb_h}
V_{\text{b}}(x) = K \cdot h(x), 
\end{equation}
where $K$ is the dielectric strength of the lubricant. Morris et al. \cite{morris2022electrical} numerically solved the electrostatics problem of the electrified EHL line contact with no discharge. They confirmed that Eq. \eqref{eq:Vb_h} is valid near the minimum film thickness where discharge is more likely to occur \cite{guo2023study,li2024study}. Several researchers have applied an elastic-electric analogy \cite{barber2003bounds,xu2024} to solve electrical contact problems. Using the elastic-electrical analogy between an elastic half-plane subjected to a point load under plane strain/stress condition \cite{johnson1987contact} and a half-plane subjected to a point current, the relation between $\widetilde{V}(x)$ and $J(x)$ can be formulated based on its counterpart relation, $h(x)$ vs. $p(x)$ in Eq. \eqref{eq:dimensional_film}, as
\begin{equation}\label{eq:VJ_relation}
\widetilde{V}(x) = K \cdot h(x) - V_0 - \frac{\rho}{2 \pi} \int_{-\infty}^{\infty} J(s) \ln |x - s|^2 \text{d}s,
\end{equation}
where $\rho = \rho_1 + \rho_2$ and $V_0$ is a constant. The current density distribution $J(x)$ must be balanced by the  current $I$ according to
\begin{equation}\label{eq:Current_equilibrium}
I = \int_{-\infty}^{\infty} J(x) \text{d}x. 
\end{equation}
Defining the following dimensionless group:

\begin{equation}\label{eq:Dimensionless_group1}
\widetilde{V}^* = \frac{R}{a^2 K} \widetilde{V}, ~~~ V_0^* = \frac{R}{a^2 K} V_0 , ~~~ J^* = \frac{\rho R}{a K} J, ~~~ I^* = \frac{\rho R}{a^2 K} I,  
\end{equation}
Eqs. \eqref{eq:electric_cont}, \eqref{eq:electric_ncont}, \eqref{eq:VJ_relation}, and \eqref{eq:Current_equilibrium} are rewritten in dimensionless forms as follows:

\begin{subequations}
\begin{align}
&\widetilde{V}^*(x^*) = 0, ~~~ J^*(x^*) > 0, ~~~  x^* \in \Omega_{\text{con}}, \label{eq:electric_cont_dimensionless} \\
&\widetilde{V}^*(x^*) > 0, ~~~ J^*(x^*) = 0, ~~~ x^* \in \Omega_{\text{ncon}}, \label{eq:electric_ncont_dimensionless} \\
&\widetilde{V}^*(x^*) = h^*(x^*) - V_0^* - \frac{1}{\pi} \int_{-\infty}^{\infty} J^*(s) \ln |x^* - s| \text{d}s, \label{eq:VJ_relation_dimensionless}\\
&I^* = \int_{-\infty}^{\infty} J^*(x^*) \text{d} x^*.
\end{align}
\end{subequations}

\subsection{Discretized equations}
Using the dimensionless domain $\Omega$ defined in Section \ref{subsec:Discrete_EHL}, the electrified lubrication interface with discharge as a linear complementarity problem (LCP) can be formulated as:

\begin{align}
&\widetilde{V}_i^* = 0, ~~~ J_i^* > 0, ~~~ i \in I_{\text{con}}, \label{eq:discrete_KKT1} \\
&\widetilde{V}_i^* > 0, ~~~ J_i^* = 0, ~~~  i \in I_{\text{ncon}}, \label{eq:discrete_KKT2} \\
&\widetilde{V}_i^* = h_i^* - V_0^* - \frac{1}{\pi} \sum_{j \in I_{\text{con}}} K_{ij}^* J_j^*, \label{eq:discrete_VJ_relation}
\end{align}
where $I_{\text{con}}$ and $I_{\text{ncon}}$ are respectively nodal index sets within nodes located at the conductive and non-conductive regions; the influence coefficient matrix $K_{ij}^*$ is the same as that in Eq. \eqref{eq:IC}. If the current is provided a priori, the LCP is constrained by 
\begin{equation}\label{eq:current_equilibrium}
I^* = \frac{\Delta_x^*}{2} \sum_{i = 1}^{n-1} \left( J_i^* + J_{i + 1}^* \right).
\end{equation}
The conjugate gradient (CG) method is used to solve Eqs. (\ref{eq:discrete_KKT1}--\ref{eq:current_equilibrium}). At the end of each iteration, the constant $V_0^*$ is adjusted as follows to ensure the current equilibrium in Eq. \eqref{eq:current_equilibrium}:
\begin{equation}\label{eq:V0_correction}
V_0^* \leftarrow V_0^* + c \left[ I^* - \frac{\Delta_x^*}{2} \sum_{i = 1}^{n-1} \left( J_i^* + J_{i + 1}^* \right) \right],
\end{equation}
where $c \in (0, 1)$ is the under-relaxation coefficient. The algorithm for the current-driven electrified EHL line contact with discharge is presented in Algorithm \ref{alg:CG1} in Appendix A. 

Due to the incompleteness of the contact and electrostatic problems in two-dimensional space, the voltage drop and the normal surface displacement are not bounded at the far end of the interface but slowly diverge toward infinity. Discharge events are mainly localized inside the contact ($|x^*| \leq 1$), and $V^*(|x^*| \in [1, R^*])$ is nearly a constant distribution. Thus, $V_R^* = V^*(x^* = -R^*)$ is chosen to represent the potential difference between two equipotential surfaces away from the discharge zone. If $V_R^*$ is prescribed, Eq. \eqref{eq:VJ_relation} can have an alternative form as follows:
\begin{equation}\label{eq:discrete_VJ_relation1}
\widetilde{V}_i^* = -V_R^* + h_i^* - \frac{1}{\pi} \sum_{j \in I_{\text{con}}} (K_{ij}^* - K_{ij}^{R*}) J_j^*,
\end{equation}
where $K_{ij}^{R^*}$ is straightly derived by setting $x_i = -R^*$ in Eq. \eqref{eq:IC}:
\begin{equation}\label{eq:IC1}
K_{ij}^{R*} = (-R^* - x_j^* + \Delta_x^*/2)( \ln|-R^* - x_j^* + \Delta_x^*/2| - 1) + (R^* + x_j^* + \Delta_x^*/2)( \ln|-R^* - x_j^* - \Delta_x^*/2| - 1).
\end{equation}
The voltage-driven electrified EHL line contact, characterized by Eqs. \eqref{eq:discrete_KKT1}, \eqref{eq:discrete_KKT2}, \eqref{eq:discrete_VJ_relation1}, is solved by the CG method, the algorithm of which is presented in Algorithm \ref{alg:CG2} in Appendix A. The current- and voltage-driven electrified EHL line contact models are implemented in Matlab. 

\subsection{Interfacial resistance and capacitance}
For modeling the electrical response of the roller bearing subjected to the alternating shaft voltage/current, each lubrication interface is simplified to a circuit composed of a resistor ($R_{\text{EHL}}$) and capacitor ($C_{\text{EHL}}$) in parallel. The dimensionless resistance of the electrified lubrication interface can be further decomposed into three resistors in parallel,
\begin{equation}
1/R_{\text{EHL}}^* = 1/R_{\text{b}}^* + 1/R_{\text{c}}^* + 1/R_{\text{d}}^*,
\end{equation}
where $R_{\text{b}}^*$ is the dimensionless bulk resistance of the lubricant, $R_{\text{c}}^*$ is the dimensionless electrical contact resistance induced by solid-solid contact, and $R_{\text{d}}^*$ is the dimensionless resistance of the plasma channel formed during dielectric breakdown of the lubricant. Since the volume resistivity of the dielectric lubricant is extremely high, the contribution of bulk resistance to $R_{\text{EHL}}^*$ is negligible. If the lubrication interface is free of the solid-solid interaction, $R_{\text{EHL}}^*$ is dominated by $R_{\text{d}}^*$,
\begin{equation}
R_{\text{EHL}}^* \approx R_{\text{d}}^* = V_R^*/I^*.
\end{equation}

\begin{figure}[h!]
  \centering
  \includegraphics[width=11cm]{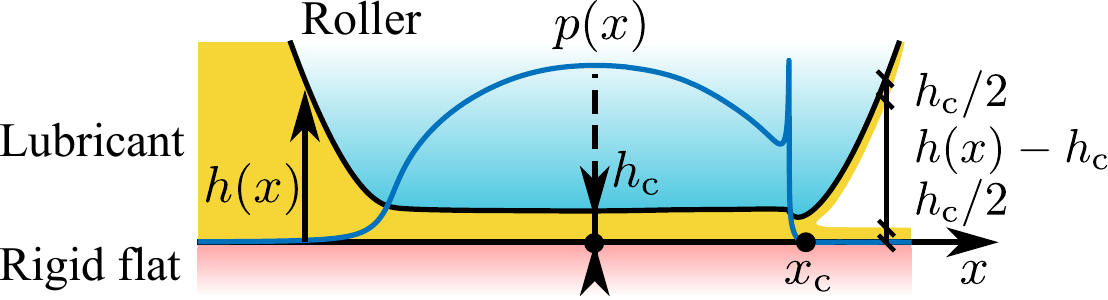}
  \caption{Schematic of the lubricant film thickness at the inlet, Hertzian contact region, and outlet}\label{fig:Fig_3}
\end{figure}

To calculate the capacitance of the lubrication interface, three assumptions are adopted: 
\begin{enumerate}
\item The relative permittivity ($\epsilon_{\text{r}}$) of the lubricant is a constant. 
\item The inlet of the EHL line contact is fully flooded within the selected computational domain \cite{dyson1965paper,ten1970influence}, as shown in Fig. \ref{fig:Fig_3}. 
\item Two surfaces form at the outlet beyond the lubricant film rupture point ($x_{\text{c}}$), where $\text{d} p/\text{d}x|_{x_{\text{c}}} = 0$, after which it assumed both surfaces are covered with thin film lubricant of thickness $h_{\text{c}}/2$ with an air gap of thickness $h(x) - h_{\text{c}}$ filling the void, see Fig. \ref{fig:Fig_3}. This assumption approximately achieves the fluid volume conservation when lubricant flows from the Hertzian contact area to the outlet \cite{dyson1965paper,ten1970influence}.  
\end{enumerate}

Based on those three assumptions, $C_{\text{EHL}} = C_{\text{ih}} + C_{\text{o}}$, where $C_{\text{ih}}$ is the capacitance of the lubricant within the inlet and Hertzian contact region before the lubricant film collapses, 
\begin{equation}
C_{\text{ih}} = \epsilon_0 \epsilon_{\text{r}} \int_{x \in \Omega_{\text{ncon}} \cap x \leq x_{\text{c}}} h^{-1}(x)\text{d}x,
\end{equation}
where $\epsilon_0$ is the permittivity of the free space, and $C_{\text{o}}$ is the capacitance of a mixture of air and lubricant in series,
\begin{equation}
C_{\text{o}} = \frac{\epsilon_0 \epsilon_{\text{r}}}{h_{\text{c}}} \int_{x > x_{\text{c}}} \left[1 + \frac{h(x) - h_{\text{c}}}{h_{\text{c}}} \epsilon_{\text{r}} \right]^{-1} \text{d}x.
\end{equation}
Defining the dimensionless capacitance as $C_{\text{EHL}}^* = C_{\text{EHL}}/(4 \pi \epsilon_0)$, 
\begin{equation}
C_{\text{EHL}}^* = \frac{\epsilon_{\text{r}} R^*}{4 \pi} \left\{ \int_{x^* \in \Omega_{\text{ncon}} \cap x^* \leq x_{\text{c}}^*} \left[h^*(x^*)\right]^{-1}\text{d}x^* + \int_{x^* > x_{\text{c}}^*} \left\{h_{\text{c}}^* + \left[h^*(x^*) - h_{\text{c}}^*\right] \epsilon_{\text{r}} \right\}^{-1} \text{d}x^* \right\}.
\end{equation}

\section{Results}

\begin{table}[h!]
\centering
\caption{Essential geometrical, mechanical, and electrical  properties of the electrified EHL line contact}
\begin{tabular}{llllll}
\toprule
Parameter & Value & Units & Parameter & Value & Units \\ 
\midrule
$G$ & $4730$ & - & $(R_1, R_2)$ & $(0.1, 0.005)$ & m \\ 
$\epsilon_{\text{r}}$  & 2.4 & - & $(x_{\text{a}}^*, x_{\text{b}}^*)$ & $(-4, 1.5)$ & - \\
$\alpha$ & $2.2 \times 10^{-8}$ & - & $z$ & 0.67 & -\\
$\eta_0$ & $0.08$ & Pa$\cdot$s& & & \\
\bottomrule
\end{tabular}\label{tab:Table_1}
\end{table}

\begin{figure}[h!]
  \centering
  \includegraphics[width=16cm]{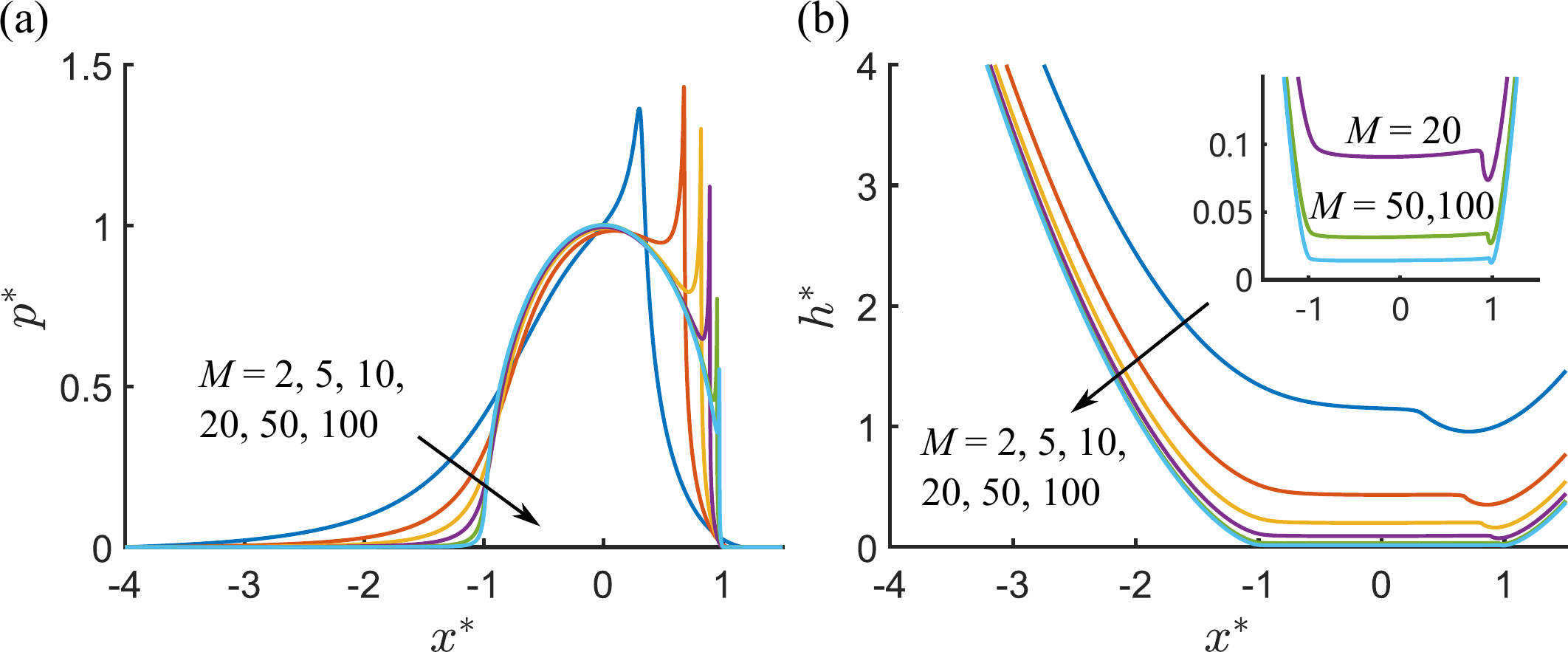}
  \caption{(a) Dimensionless pressure and (b) dimensionless film thickness distributions with various Moe's parameters: $M = W(2U)^{-1/2} \in [2, 100]$ and $L = G (2U)^{1/4}= 10$. Other essential parameters are given in Table \ref{tab:Table_1}.}\label{fig:Fig_R1}
\end{figure}

The dimensionless Reynolds equation is solved by the multigrid method with $3585$ and $15$ nodes on the finest and coarsest levels, respectively. The converged solutions ($p^*$ and $h^*$) over the finest level after $10$ W-cycles with various Moe's parameters $(M, L)$ are shown in Fig. \ref{fig:Fig_R1}(a,b). The Pan-Hamrock minimum film thickness formula \cite{Pan89, hamrock2004fundamentals}, 
\[
h_{\text{m}}^* = 1.714 W^{-0.128} U^{0.694} G^{0.568} (R/a)^2,
\] 
is used to support the validity of the present EHL solver, with mean and maximum relative differences of $5.42 \%$ and $9.24 \%$, respectively. 

\begin{figure}[h!]
  \centering
  \includegraphics[width=16cm]{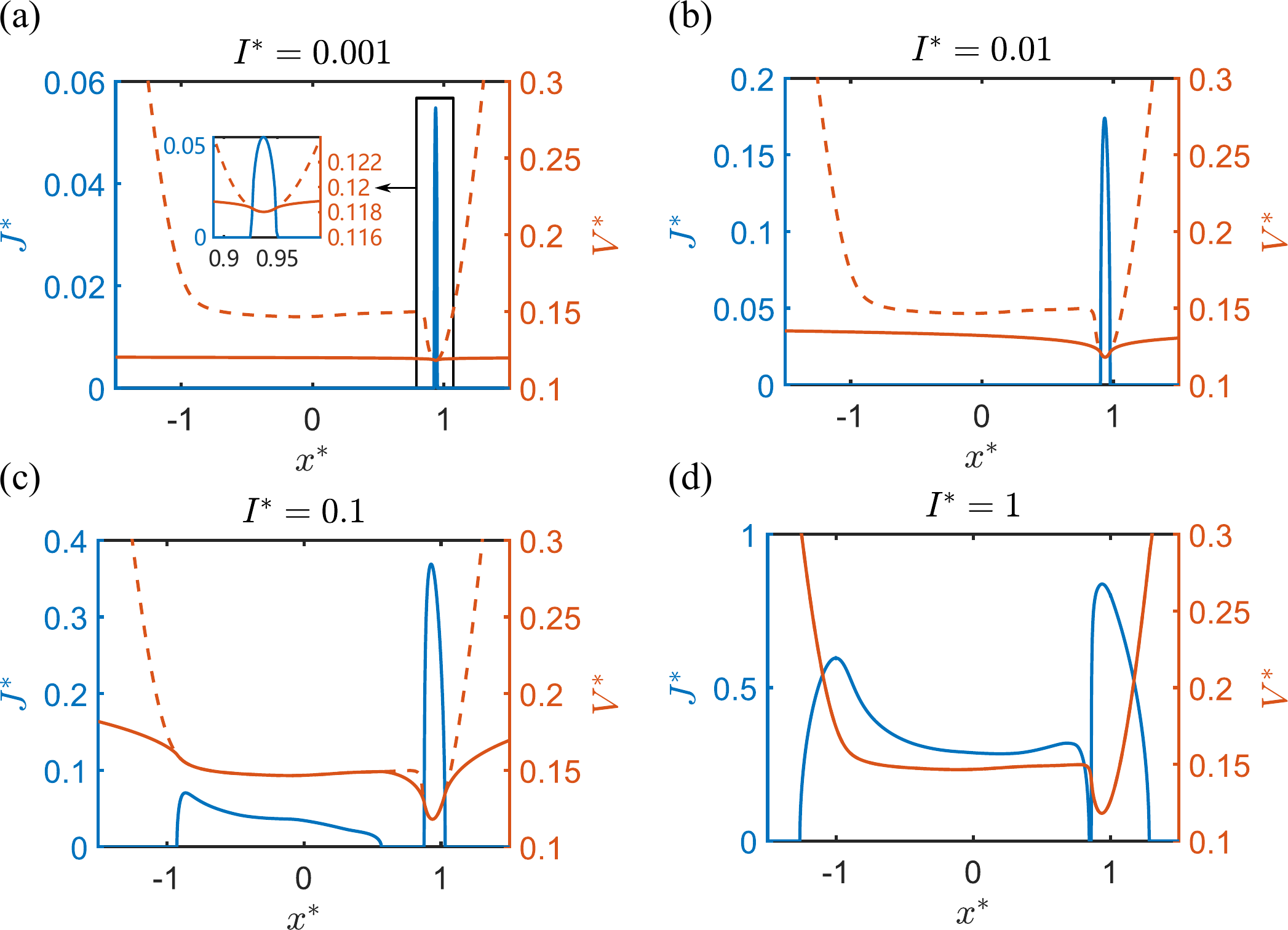}
  \caption{Dimensionless current density ($J^*$) and voltage drop ($V^*$) distributions with four different values of $I^*$: (a) 0.001, (b) 0.01, (c) 0.1, (d) 1. $M = 20$, $L = 10$. The dashed line corresponds to the dimensionless breakdown voltage, which is the same as $h^*$. Other essential parameters are given in Table \ref{tab:Table_1}.}\label{fig:Fig_R2}
\end{figure}

The electrical properties of dimensionless current density ($J^*$) and voltage drop ($V^*$) at the lubricated interface subjected to an increasing dimensionless current $I^*$ across the contact are shown in Fig. \ref{fig:Fig_R2}. When, the current across the contact is low, $I^* = 0.001$, the non-zero $J^*$ occurs only at the minimum film thickness location where the breakdown voltage is exceeded (dashed line in Fig. \ref{fig:Fig_R2}(a)). According to Eq. \eqref{eq:VJ_relation_dimensionless}, the dimensionless breakdown voltage is the same as the dimensionless film thickness (i.e., $V_{\text{b}}^* = h^*$) and is invariant of dimensionless current $I^*$. Therefore, Fig. \ref{fig:Fig_R2}(a) supports the experimental observation that discharge occurs at the minimum film thickness \cite{sunahara2008development,sunahara2011preliminary,guo2023study}. Within the discharge zone, the potential drop is forced to conform to the breakdown voltage, as shown in the inset of Fig. \ref{fig:Fig_R2}(a). The current density exhibits a parabolic shape, which is nearly symmetric about the location of the minimum film thickness. Since current density $J^*$ occupies a localised discharge zone, its effect on $V^*$ rapidly vanishes spatially so that $V^*$ remains constant within the investigated ranges outside the discharge zone. The dimensionless current density, $J^*$, remains zero outside the discharge zone. As $I^*$ increases to $0.01$ (see Fig. \ref{fig:Fig_R2}(b)), the discharge zone expands. The potential drop is perturbed locally by $J^*$ within the discharge zone. The difference between $V^*$ and $V_{\text{b}}^*$ over the Hertzian contact area excluding the discharge zone reduces, indicating that a further increase of $I^*$ can lead to nucleation of new discharge zones. This expectation is confirmed in Fig. \ref{fig:Fig_R2}(c), where the current flow ($I^*$) is increased to $0.1$. A new discharge zone is nucleated at the inlet within the Hertzian contact area. The potential drop, $V^*$, is non-uniform over the investigated interface, and the current density ($J^*$) is distributed across two discrete regions. When $I^* = 1$, the discharge zone exceeds the Hertzian contact area. The difference between $V^*$ and $V_{\text{b}}^*$ is almost non-distinguishable. The current density $J^*$ is nearly continuously distributed across the Hertzian contact area. Fig. \ref{fig:Fig_R2} indicates that, for higher currents to pass across the EHL film an increase in the extent of the contact across which dielectric breakdown occurs must take place. During these instances the electrified EHL line contact transits from a capacitive type to a resistive type. 

\begin{figure}[h!]
  \centering
  \includegraphics[width=16cm]{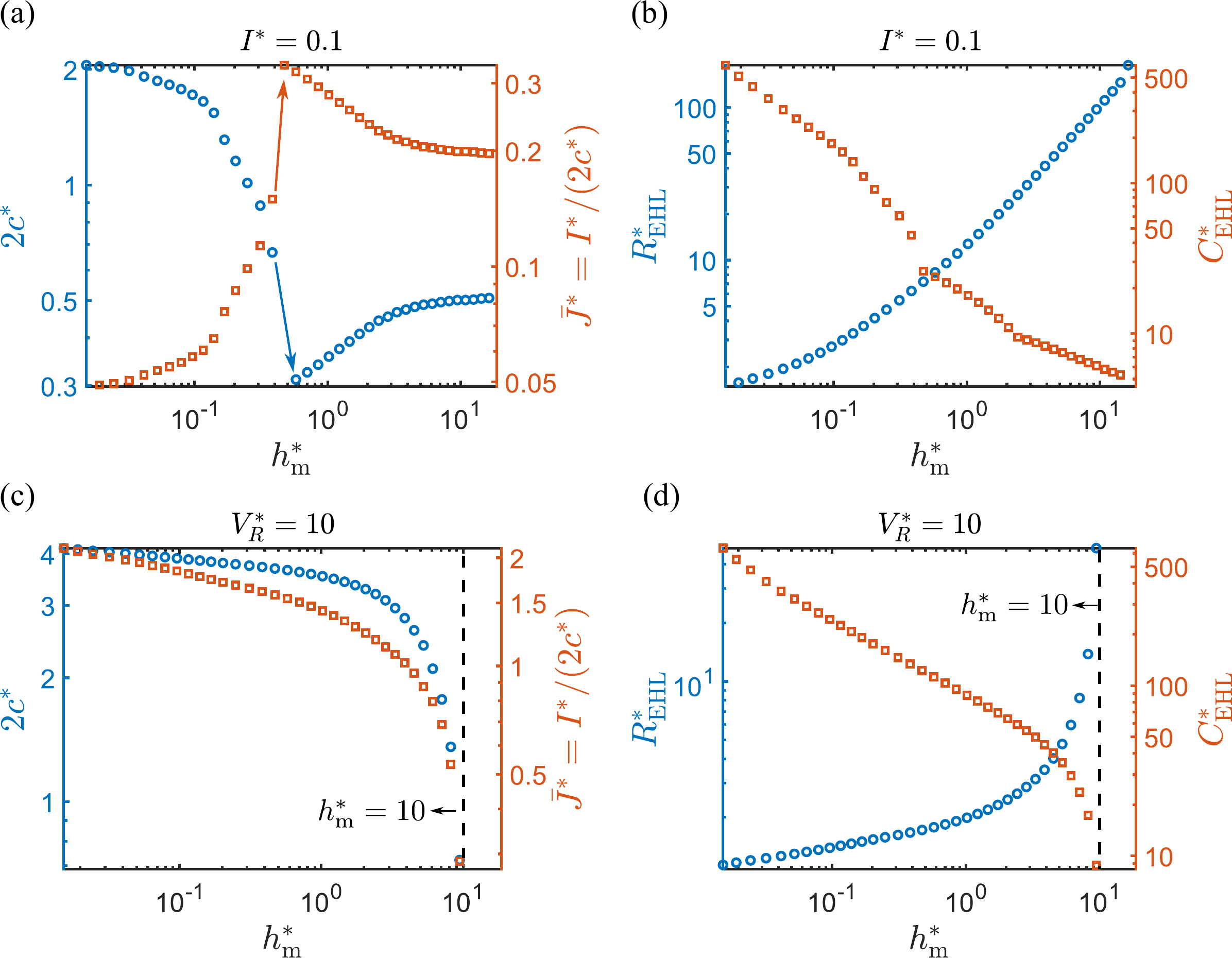}
  \caption{Variations of the dimensionless width of the discharge zone and electrical discharge resistance with the minimum film thickness: (a) current-driven condition with $I^* = 0.1$; (b) voltage-driven condition with $V_{\text{R}}^* = 10$. Other essential parameters are given in Table \ref{tab:Table_1}.}\label{fig:Fig_R3}
\end{figure}

\begin{figure}[h!]
  \centering
  \includegraphics[width=16cm]{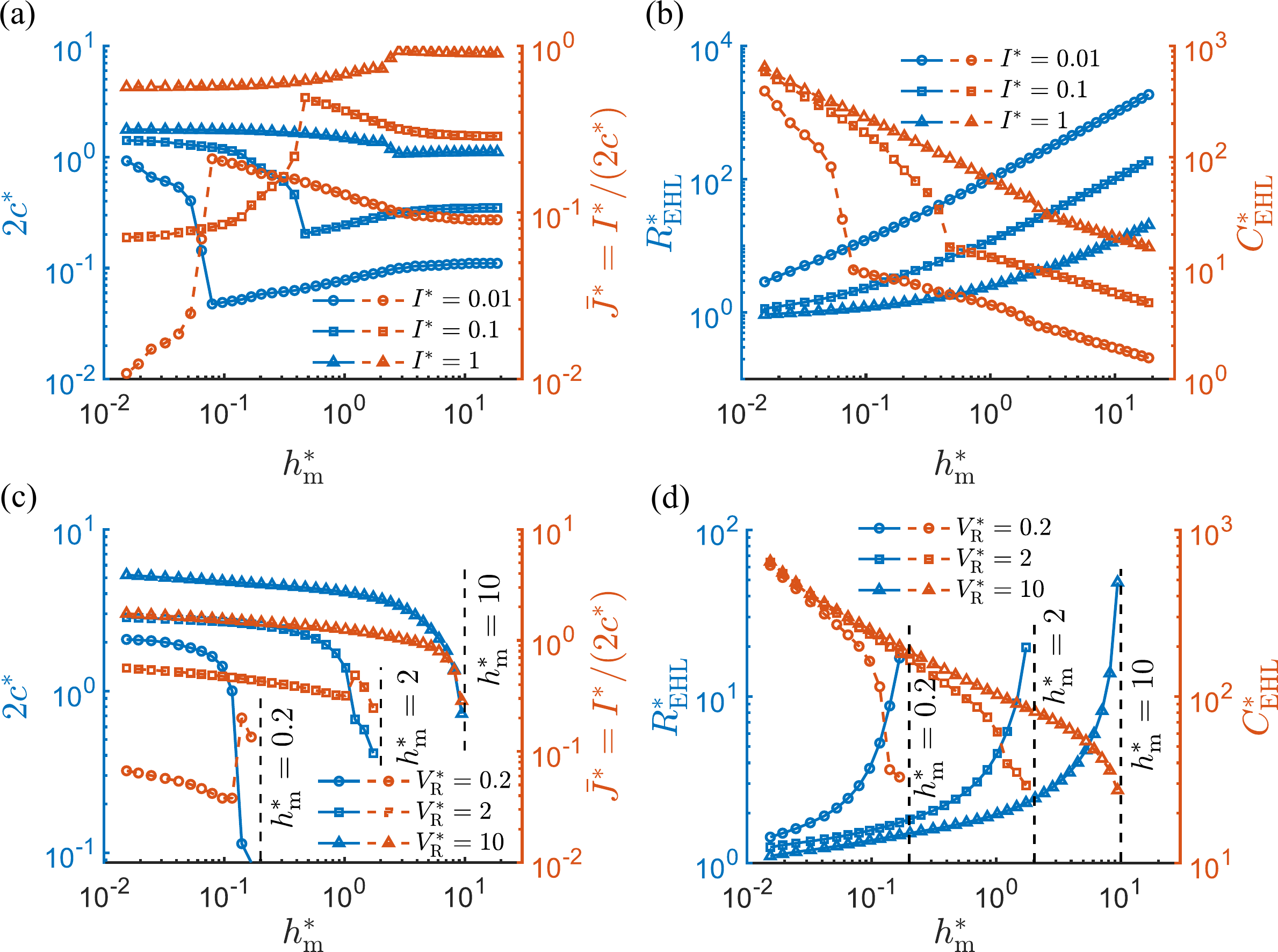}
  \caption{Variation of discharge width ($2c^*$), mean current density ($\bar{J}^*$), EHL interfacial resistance ($R_{\text{EHL}}^*$), and capacitance ($C_{\text{EHL}}^*$) with $h_{\text{m}}^*$ under (a,b) current-driven condition with $I^* = 0.01, 0.1, 1$, and (c,d) voltage-driven condition with $V_R^* = 0.2, 2, 10$.}\label{fig:Fig_R4}
\end{figure}

Besides the applied electrical load, the electrical properties at the lubrication interface can be manipulated by changing the minimum film thickness, $h_{\text{m}}^*$, as shown in Fig. \ref{fig:Fig_R3} and \ref{fig:Fig_R4}. Under the current-driven condition, the dimensionless discharge width (denoted by $2c^*$, where $c^*$ is the semi-width of the discharge zone) first decreases and then increases with a decreasing $h_{\text{m}}^*$, see Fig. \ref{fig:Fig_R3}(a). The mean current density, $\bar{J}^* = I^*/(2 c^*)$, shows the inverse trend. The transition points of the non-monotonic trends of $2c^*$ and $\bar{J}^*$ are due to the nucleation of the second discharge zone at the inlet. The resistance and capacitance of the EHL line contact ($R_{\text{EHL}}^*$ and $C_{\text{EHL}}^*$) under the current-driven condition increase and decrease, respectively, as $h_{\text{m}}^*$ increases, see Fig. \ref{fig:Fig_R3}(b). The capacitance relation, $C_{\text{EHL}}^*(h_{\text{m}}^*)$, shows $C_1$ discontinuity at the transition point of the non-monotonic trend of $2c^*$, while $R_{\text{EHL}}^*(h_{\text{m}}^*)$ remains smooth. The voltage-driven condition is more likely to occur in practice, and the discharge only occurs when $V_{\text{R}}^* > h_{\text{m}}^*$ since $h_{\text{m}}^* = \min\left( V_{\text{b}}^*(x^*) \right)$. When $V_{\text{R}}^* > h_{\text{m}}^*$, numerical solutions diverge. Unlike the current-driven condition, $2c^*$ and $\bar{J}^*$, under the voltage-driven condition, monotonically drop to zero as $h_{\text{m}}^*$ increases toward $h_{\text{m}}^* = 10$, see Fig. \ref{fig:Fig_R3}(c). Fig. \ref{fig:Fig_R3}(d) shows a similar monotonic variations of $R_{\text{EHL}}^*$ and $C_{\text{EHL}}^*$ with $h_{\text{m}}^*$ as that in Fig. \ref{fig:Fig_R3}, where the resistance and capacitance, respectively, become infinite and a finite value when $h_m^* \to 10$. Fig. \ref{fig:Fig_R3}(b, d) indicates that the discharge transits from the resistive type to the capacitive type as $h_{\text{m}}^*$ increases. Under the voltage-driven condition, the open circuit status is reached once $h_{\text{m}}^* \geq 10$.  

The applied electrical loads strongly influence the electrical properties of the electrified EHL line contact. Under the current-driven condition, the discharge width ($2 c^*$) and mean current density ($\bar{J}^*$) increases as $I^*$ increases. The abrupt transition point gradually shifts toward the lower minimum film thickness as $I^*$ reduces. The electrical resistance ($R_{\text{EHL}}^*$) and capacitance ($C_{\text{EHL}}^*$) reduces and increases, respectively, with the increase of $I^*$. As $h_{\text{m}}^*$ further reduces, it is expected that $2c^*$, $R_{\text{EHL}}^*$ and $C_{\text{EHL}}^*$ gradually converge to the corresponding solutions of electrical contact between two solid electrodes. Under the voltage-driven condition, the electrical properties behave similarly to those driven by the current. At $V_{\text{R}}^* = 0.2$ and $2$, a second peak of mean current density appears before $\bar{J}^*$ reduces to zero. As $V_{\text{R}}^*$ reduces, the electrified line contact becomes open-circuit at a lower minimum film thickness. 

\section{Discussion}

\subsection{Influence of roughness}
\begin{figure}[h!]
  \centering
  \includegraphics[width=16cm]{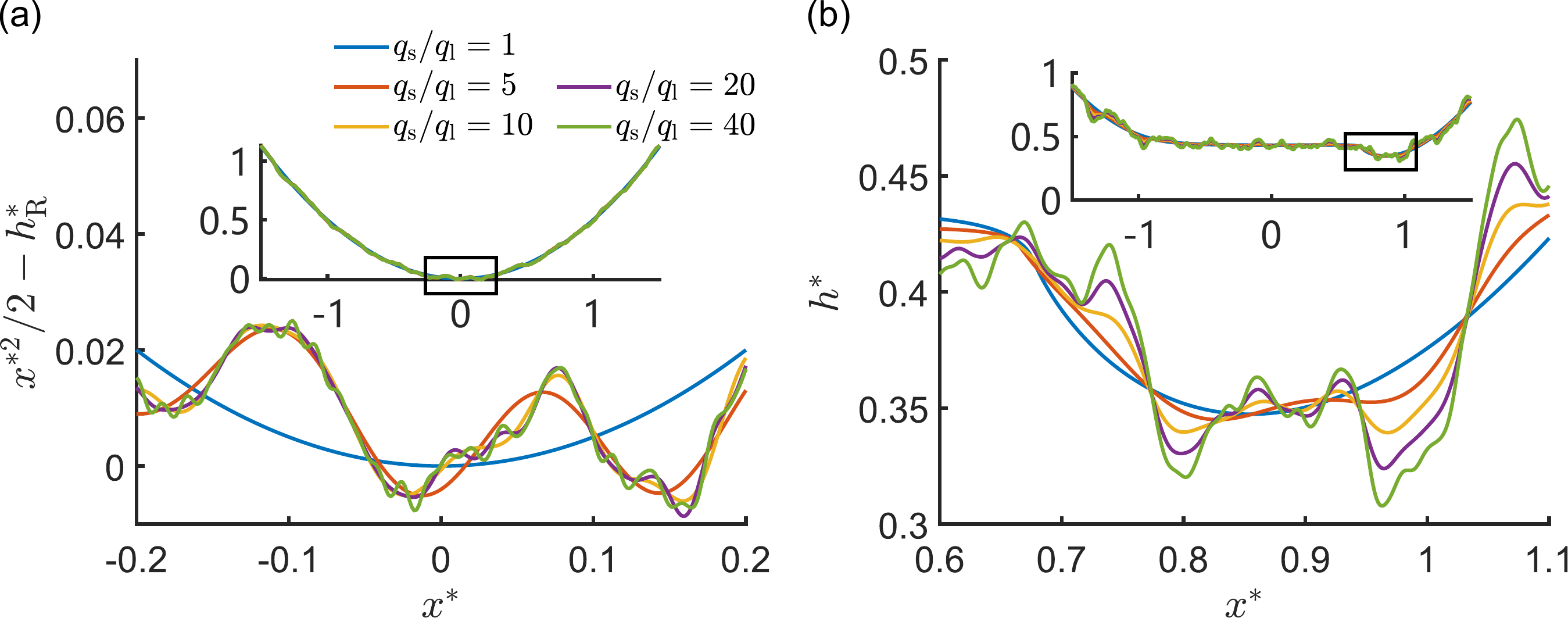}
  \caption{(a) Undeformed rough profile of roller with various values of upper cut-off wavenumber at a small vicinity of the center of EHL line contact; Inset: global view of the undeformed rough profile of roller; (b) Deformed rough profile of roller with various values of upper cut-off wavenumber at the minimum film thickness location; Inset: global view of the deformed rough profile of roller. Other essential parameters besides those listed in Table 1 are: $M = 5$, $R_{\text{q}} = h_{\text{m}}/6$ for $q_{\text{s}}/q_{\text{l}} = 5$, $H = 0.8$, $\lambda_{\text{l}} = (x_{\text{b}} - x_{\text{a}})/10$. }\label{fig:Fig_R5}
\end{figure}
The EHL contact, in practice, is inevitably formed by rough surfaces. Four random, self-affine, bandwidth-limited, rough profiles with topographies, $h_{\text{R}}(x)$, are generated and superposing it to the undeformed, parabolic profile, $h(x) = x^2/2R$. All rough profiles share a similar power spectral density as follows:  
\begin{equation}
C(q) = 
\begin{cases}
C_0 \cdot q^{-2H - 1} ~~~ q \in [q_{\text{l}}, q_{\text{s}}], \\
0 ~~~ \text{elsewhere},
\end{cases}
\end{equation}
where $q$ is the angular wavenumber, which is $2 \pi$ divided by the wavelength, $q_{\text{l}} = 2\pi/\lambda_{\text{l}}$ and $q_{\text{s}} = 2\pi/\lambda_{\text{s}}$, are lower and upper cut-off wavenumbers associated with the longest and shortest wavelengths $(\lambda_{\text{l}}, \lambda_{\text{s}})$, $H \in (0, 1)$ is the Hurst exponent, and $C_0$ is a constant. The rough profiles are numerically generated through the inverse Fourier transform of the magnitude spectrums, which are constructed based on $C(q)$ with four different ratios of $q_{\text{s}}/q_{\text{l}}$ and the same random phase array, see Fig. \ref{fig:Fig_R5}(a) for a graphical illustration of the rough roller superposed with various rough profiles. The four generated rough profiles have nearly identical root mean square roughness, $R_{\text{q}} = \sqrt{\langle |h_{\text{R}}|^2 \rangle} \approx 13$ nm. The deformed surface profile with and without roughness are shown in Fig. \ref{fig:Fig_R5}(b), where the profile with higher $q_{\text{s}}/q_{\text{l}}$ has a larger perturbation of the film thickness and sharper asperities (larger peak curvatures). 

\begin{figure}[h!]
  \centering
  \includegraphics[width=13cm]{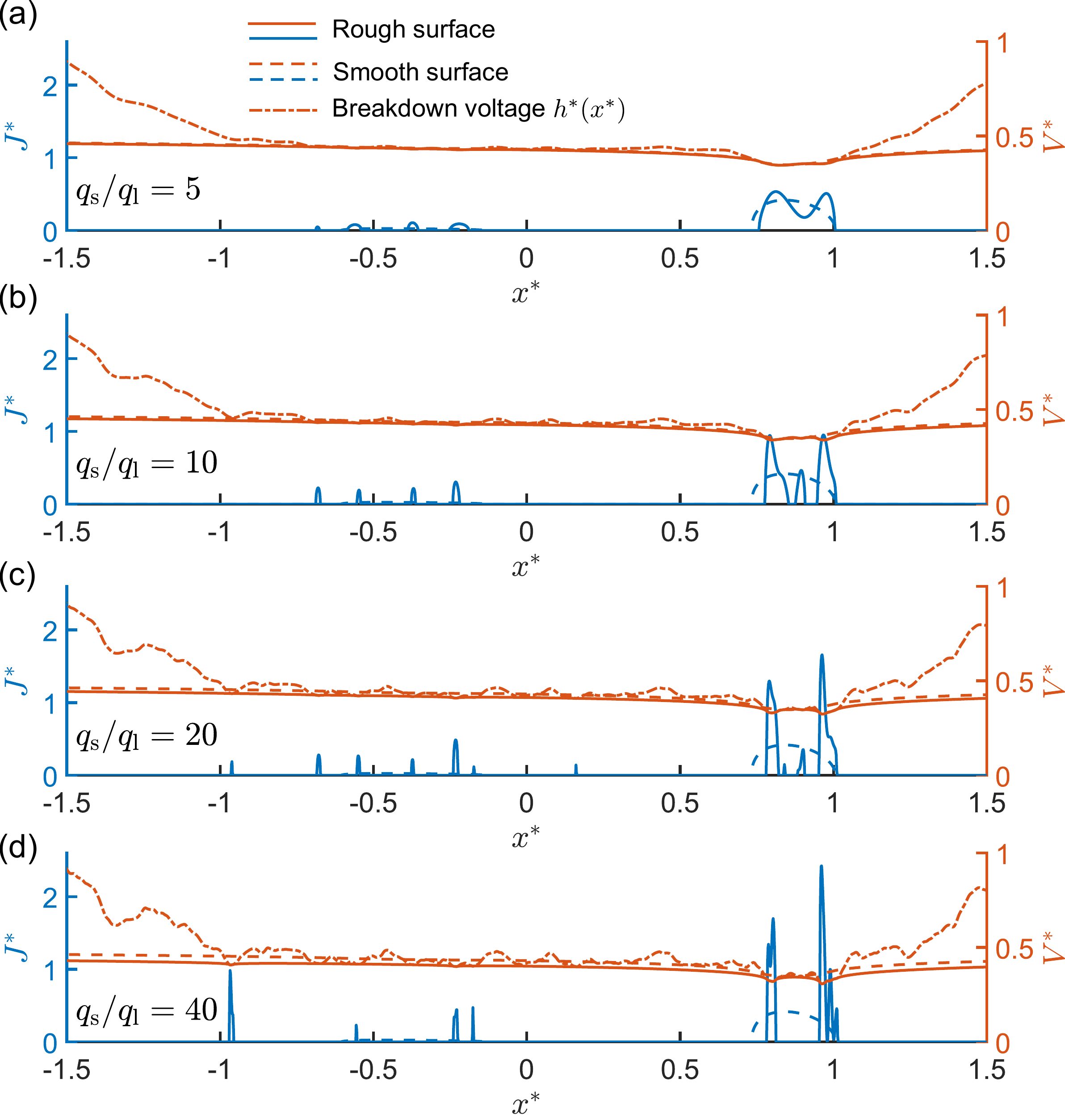}
  \caption{Distributions of the dimensionless current density and dimensionless voltage over the rough EHL line contact interface. Four different values of upper cut-off wavenumber are used: $q_{\text{s}}/q_{\text{l}} =$ (a) $5$, (b) $10$, (c) $20$, (d) $40$. The deformed rough profile is taken from Fig. \ref{fig:Fig_R5}(b). The solid and dashed lines represent $J^*$ and $V^*$ of rough and smooth EHL line contact, respectively. The dash-dotted line corresponds to the breakdown voltage. The limits of the y-axis of (a-d) are kept the same. $I^* = 0.1$.
}\label{fig:Fig_R6}
\end{figure}

The distributions of current density ($J^*$) and voltage drop ($V^*$) associated with rough and smooth profiles are shown in Fig. \ref{fig:Fig_R6}, where the potential drop (red solid line) is strictly lower than or equal to the breakdown voltage (dash-dotted line), indicating the robustness of the numerical algorithm for various rough electrified EHL line contact problems. The potential drop distribution is not perturbed by the roughness. The current density distribution at the minimum film thickness location transits from a parabolic shape ($q_{\text{s}}/q_{\text{l}} = 1$) to a wavy profile ($q_{\text{s}}/q_{\text{l}} = 5$). Multiple current spikes are formed in the central film thickness region due to the roughness-induced corona discharge at asperity peaks and split into multiple distinct discharge sub-regions when $q_{\text{s}}/q_{\text{l}} > 5$ increases, see Fig. \ref{fig:Fig_R6}(b-d). As $q_{\text{s}}/q_{\text{l}}$ increases from $5$ to $40$, the mean absolute curvature, $\langle |\partial h_{\text{R}}/\partial x| \rangle$, increases from $1667$ to $19070$ monotonically by more than $10$ times. According to the deformed rough profile in Fig. \ref{fig:Fig_R5}(b), discharge sub-regions are centered approximately about asperity peaks with sharp tips, see Fig. \ref{fig:Fig_R5}(b). It is also important to note as $q_{\text{s}}/q_{\text{l}}$ increases from $5$ to $40$ the mean dimensionless current density over discharge zones increases from $0.13$ to $0.81$. Current density has been strongly correlated with increased bearing damage as result of an increase in localised Joule heating at the surface \cite{busse1997bearing}. As the result the introduction of roughness is shown to exacerbate bearing damage. 

\begin{figure}[h!]
  \centering
  \includegraphics[width=16cm]{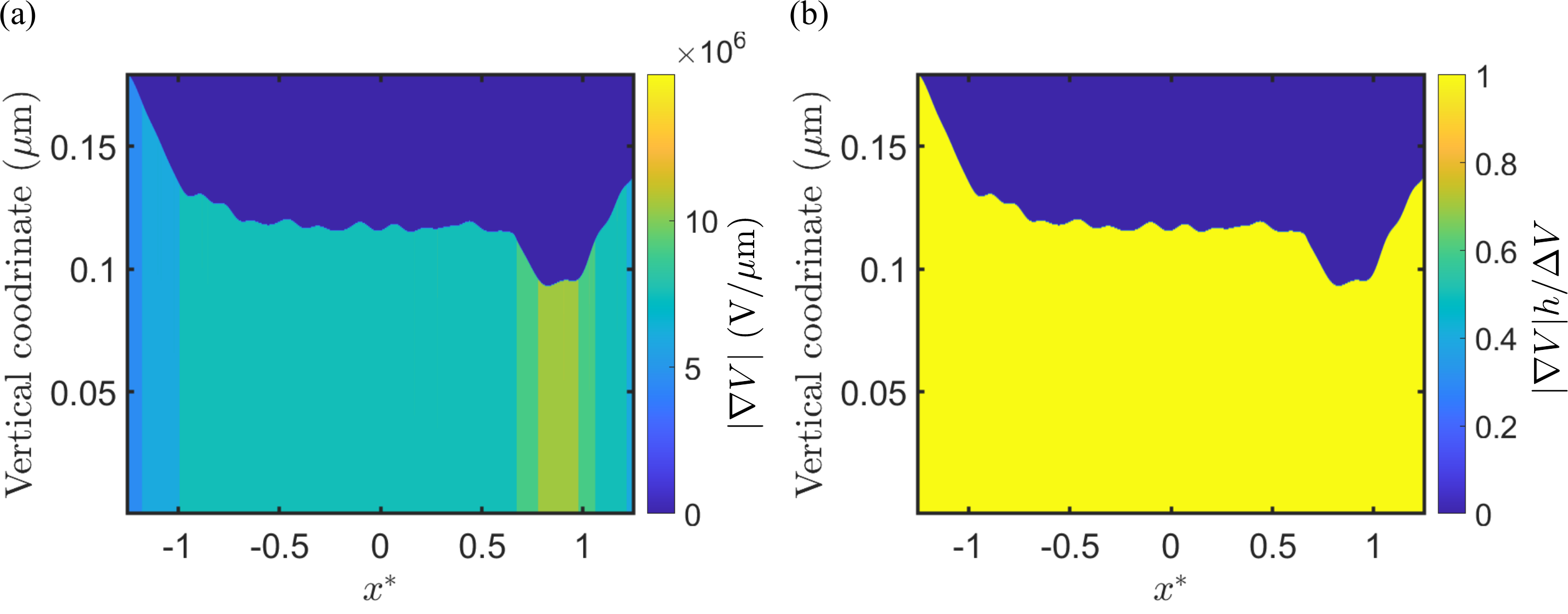}
  \caption{(a) Distribution of the dimensional magnitude of the electric field vector ($|{\bf E}|$) without discharge; (b) Distribution of the dimensionless electric field ($|{\bf E}|h/\Delta V$). All the essential parameters are the same as those in Figs. \ref{fig:Fig_R5} and \ref{fig:Fig_R6} with $q_{\text{s}}/q_{\text{l}} = 5$. The voltage drop across the lubricant is $\Delta V = 1$ V.}\label{fig:Fig_R7}
\end{figure}

\subsection{Validity of parallel-plate capacitor theory}
In the present model, the discharge occurs when $V(x)/h(x) = K$. This criteria implicitly assumes that the vertical component of the electric field is uniform along the thickness  ($z$) direction and the tangential component is negligible. To test the validity of this criteria, it is assumed that the discharge does not occur in the lubricant, and the following Laplace's equation is solved using the finite difference method developed by Morris et al. \cite{morris2022electrical}: 
\begin{equation}
\partial^2 V/\partial x^2 + \partial^2 V/\partial z^2 = 0 
\end{equation}
where the $z$-axis has its origin at the rigid flat and points upwardly toward the roller. The boundary conditions are $V(x, y = h(x)) = \Delta V$ and $V(x, y = 0) = 0$. The electric field vector is given by ${\bf E} = -\nabla V$. The five electrified, lubricated rough interfaces solved in Fig. \ref{fig:Fig_R6} are investigated again. The distribution of the magnitude of the electric field vector $|\bf E|$ over the lubricant and roller is given in Fig. \ref{fig:Fig_R7}(a), where $q_{\text{s}}/q_{\text{l}} = 5$. The electric field has the maximum magnitude at the minimum film thickness location. As the lubricant film increases, the magnitude monotonically increases. The roller is an equipotential surface with zero electric field inside. The distribution of the ratio of $|\nabla V|$ to $\Delta V/h(x)$ is shown in Fig. \ref{fig:Fig_R7}(b). The mean value inside the lubricant is $1.0088$, which is a strong signal that the tangential component of the electric field, $E_x = -\partial V/\partial x$, is negligible. The same conclusion can be drawn for all other cases investigated in Fig. \ref{fig:Fig_R6}. 

\subsection{Limitations of the present model}
There are several limitations in the present numerical model that can be improved in future work: (1) the dielectric constant is uniform over the lubricant thin film so that the discharge is independent of the dielectric constant. According to the Clausius-Mossotti equation \cite{morris2022electrical}, it depends on the local density of the lubricant. Thus, the discharge should be strongly correlated with the local pressure and temperature; (2) The electrified interface solved in the present work is subjected to a DC voltage or current. In the bearings of the electric motor, the lubrication interface is subjected to an alternating current or voltage with a high frequency (up to 10k Hz). Additionally, the present model solves an idealized electrified lubrication interface with zero electrical loads. In reality, the electrified lubricated interface is connected with other resistance and capacitance due to the other electrified interface and digital meters. For an electrified roller element bearing, a complex RC circuit contributed by all electrified interfaces, rollers, inner and outer rings, and cage should be formulated. This RC circuit should be solved first in order to get the output voltage applied across the interface, which is essentially the boundary condition of the present work. (3) The present work is efficient in solving one electrified interface. However, the computational time of simulating the electrical response of an electrified roller bearing becomes impractical because of a large amount of lubrication interface and short time step used to account for the high frequency alternating shaft voltage or current. The analytical model for characterizing the electrical response of a single lubrication interface is paramount. (4) Electrical conductivity of the lubricant is extremely low, typically in the range of $[1 \times 10^{-11}, 1 \times 10^{-10}]$ S/cm at $30^{\circ}$C \cite{mcfadden2016electrical}, which results in nearly zero current passing through the non-discharge zone. Therefore, it was not included in this study. Increasing the electrical conductivity of the lubricated interface (e.g., via conductive nanoparticles) or promoting mixed lubrication (by enhancing asperity contact) can enable greater current flow through the non-discharging zone. This reduces the current density over the pitting area, thereby mitigating EIBD in electrified bearings. The electrical conductivity of the lubricant and interface should be considered in future studies, particularly in mixed or boundary lubrication regimes. (5) Joule heating introduced by the applied electric field was shown to reduce the film thickness \cite{li2024study,liu2025evolution,li2024studya} (by approximately $50\%$) at high voltages. However, for contact potentials that are less than two times the breakdown voltage, the film thickness was shown to be unaffected by Joule heating \cite{liu2025evolution}. In this study, we limit the analysis to these low-voltage conditions, which allow one-way coupling of EHL to electrical contact. Driven by the need to move toward electric powertrains with higher voltage architectures, a full coupling between EHL and electrical contact should be considered in future studies. Finally, as nearly all electrified ball-on-disk tests have focused on the rolling element instead of the cylindrical roller, it is an important future step to extend the current work to the point EHL contact, enabling a full experimental validation of the present discharge model with point contact EIBD test results.

\section{Conclusion}
In the present work,  a novel numerical model for solving the discharge problem in the electrified EHL line contact interface is presented. The numerical results demonstrate that, as the discharge current increases, the location of the discharge zone transits from the neighborhood of the minimum film thickness to the entire Hertzian contact zone and beyond. The electrical properties of the discharge interface (including the width of the discharge zone, mean current density within the discharge zone, resistance, and capacitance of the lubricated interface) are strongly dependent on the applied electrical loads across the interface and the minimum film thickness. As higher frequency roughness components are superposed onto the existing topography, current density distribution transits from concentrated discharge at the neighborhood of the minimum film thickness to multiple point discharge at discrete locations with higher density. In addition, current density is shown to increase with the introduction of roughness and particularly with roughness profiles with high peak curvatures. Future work should focus on including the density-dependent dielectric constant and alternating current/voltage in the present numerical model. Additionally, the analytical model for characterizing the electrical response of an EHL line contact should be derived for the future circuit analysis of the electrified roller bearing. In conclusion, our study sheds light on the electrical properties of the inaccessible lubrication interface with discharge. This numerical model will pave the way for future research on relieving or even permanently solving EIBD problems in electric motor bearings.

\section*{Acknowledgments}
This work was supported by the Fundamental Research Funds for the Central Universities (No. PA2024GDSK0044, JZ2025HGTG0298). 

\begin{spacing}{1} 
	\bibliographystyle{asmejour}
	\bibliography{ref}
\end{spacing}
%
\setcounter{figure}{0}
\setcounter{equation}{0}
\renewcommand{\thefigure}{A.\arabic{figure}}
\renewcommand{\theequation}{A.\arabic{equation}}

\section*{Appendix A. Algorithms of EHL line contact with discharge}

\begin{algorithm}[h!]\label{alg:CG1}
\setstretch{1}
\SetKwInOut{Input}{input}
\SetKwInOut{Output}{output}
\caption{Electrified EHL line contact with discharge event under current-driven condition \\
The algorithm solves Eqs. (\ref{eq:discrete_KKT1}--\ref{eq:current_equilibrium}) using the CG method developed by Polonsky and Keer \cite{polonsky1999numerical} with minor changes. The boldsymbol represents the column vector, except for the influence matrix $\boldsymbol{K}^*$ which is composed of the dimensionless influence coefficient $K_{ij}^*$. $\epsilon_0 = 1 \times 10^{-6}$ is suggested.}
\Input{$\boldsymbol{x^*}$, $ \boldsymbol{h^*}$, $\boldsymbol{K}^*$, $\boldsymbol{K}^{R*}$, $\epsilon_0$, $I^*$.}
\Output{$\boldsymbol{J}^*$, $\boldsymbol{V}^*$, $V_R^*$.}
Initialize $\epsilon_{\text{p}} = \epsilon_{\text{I}} = 1$, $I_{\text{con}} = \{i |x_i^* \in [-1, 1]\}$\;
Initialize $J_{i \in I_{\text{con}}}^* = I^*/2$, $J_{i \notin I_{\text{con}}}^* = 0$, $\boldsymbol{J}_{\text{old}}^{*} =  \boldsymbol{J}^*$\tcp*{Current equilibrium}
Initialize $V_0^* = \text{mean} \left( h_{i \in I_{\text{con}}}^* - \frac{1}{\pi} \left( \boldsymbol{K}^* \boldsymbol{J}^* \right)_{i \in I_{\text{con}}} \right)$\tcp*{Zero voltage drop at $I_{\text{con}}$}

\While{$\epsilon_{\text{p}} > \epsilon_0$ \textbf{or} $\epsilon_{\text{I}} > \epsilon_0$}{
  $\widetilde{\boldsymbol{V}}^* = \boldsymbol{h}^*- V_0^* - \frac{1}{\pi} \boldsymbol{K}^* \boldsymbol{J}^*$ \tcp*{Eq.\eqref{eq:discrete_VJ_relation}}
  $\boldsymbol{t} = \text{zeros}(n, 1)$\;
  $t_{i \in I_{\text{con}}} = \widetilde{V}_{i \in I_{\text{con}}}^*$ \tcp*{Calculate the conjugate direction}
  $\boldsymbol{r}' = -\frac{1}{\pi} \boldsymbol{K}^*\boldsymbol{t}$\;
  $\tau = \sum \limits_{i \in I_{\text{con}}} \widetilde{V}_i^* t_i/\sum \limits_{i \in I_{\text{con}}} r'_i t_i$ \tcp*{Calculate the step size}

  $\boldsymbol{J}^* \leftarrow \boldsymbol{J}^* - \tau {\bf t}$ \tcp*{Correct $\boldsymbol{J}^*$ along the conjugate direction}

  {\bf Set} all negative $J_i^*$ to zero\;

  {\bf Update} $I_{\text{ol}} = \{i|J_i^* = 0, \widetilde{V}_i^* < 0, i = 1, \cdots, n \}$\;

  $J_{i \in I_{\text{ol}}}^* \leftarrow J_{i \in I_{\text{ol}}}^* - \tau \widetilde{V}_{i \in I_{\text{ol}}}^*$\tcp*{Correct negative $\widetilde{V}_i^*$}
  
  $V_0^* \leftarrow V_0^* + c \left[ I^* - \frac{\Delta_x^*}{2} \sum\limits_{i = 1}^{n-1} \left( J_i^* + J_{i + 1}^* \right) \right]$ \tcp*{Correct $V_0^*$ based on Eq.\eqref{eq:V0_correction}}
    
  $\epsilon_{\text{p}} = \sqrt{\sum \limits_{i = 1}^n (J_i^* - J_i^{^*\text {old}})^2}/\sqrt{\sum \limits_{i = 1}^n J_i^{*2}}$ \tcp*{Error}
  
  $\epsilon_{\text{I}} = \left[I^* - \frac{\Delta_x^*}{2} \sum \limits_{i = 1}^{n-1} \left( J_i^* + J_{i + 1}^*\right)\right]/I^*$
  
  $\boldsymbol{J}^{*\text{old}} = \boldsymbol{J}^*$, $I_{\text{con}} = \{ i |J_i^* > 0, i = 1, \cdots, n \}$\;
 }
$\boldsymbol{V}^* = \boldsymbol{h}^* - \widetilde{\boldsymbol{V}}^* $\; 
$V_R^* = \frac{\pi}{2} \boldsymbol{K}^{R*} \boldsymbol{J} + V_0^*$\;
\end{algorithm}

\begin{algorithm}[h!]\label{alg:CG2}
\setstretch{1}
\SetKwInOut{Input}{input}
\SetKwInOut{Output}{output}
\caption{Electrified EHL line contact with discharge events under voltage-driven condition\\
The algorithm solves Eqs. \eqref{eq:discrete_KKT1}, \eqref{eq:discrete_KKT2}, and \eqref{eq:discrete_VJ_relation1} using the CG method developed by Polonsky and Keer \cite{polonsky1999numerical} with minor changes. The boldsymbol represents the column vector, except for the influence matrix $\boldsymbol{K}^*$ and $\boldsymbol{K}^{R*}$ which are composed of the dimensionless influence coefficients $K_{ij}^*$ and $K_{ij}^{R*}$, respectively. $\epsilon_0 = 1 \times 10^{-6}$ is suggested.}
\Input{$\boldsymbol{x^*}$, $ \boldsymbol{h^*}$, $\boldsymbol{K}^*$, $\boldsymbol{K}^{R*}$, $\epsilon_0$, $V_R^*$.}
\Output{$\boldsymbol{J}^*$, $\boldsymbol{V}^*$, $I^*$.}
Initialize $\epsilon_{\text{p}} = 1$, $I_{\text{con}} = \{i |x_i^* \in [-1, 1]\}$\;
Initialize $J_{i \in I_{\text{con}}}^* = V_R^*$, $J_{i \notin I_{\text{con}}}^* = 0$, $\boldsymbol{J}_{\text{old}}^{*} =  \boldsymbol{J}^*$\tcp*{Current equilibrium}

\While{$\epsilon_{\text{p}} > \epsilon_0$}{
  $\widetilde{\boldsymbol{V}}^* = -V_R^* + \boldsymbol{h}^* - \frac{1}{\pi} (\boldsymbol{K}^* - \boldsymbol{K}^{R*}) \boldsymbol{J}^*$ \tcp*{Eq.\eqref{eq:discrete_VJ_relation1}}
  $\boldsymbol{t} = \text{zeros}(n, 1)$\;
  $t_{i \in I_{\text{con}}} = \widetilde{V}_{i \in I_{\text{con}}}^*$ \tcp*{Calculate the conjugate direction}
  $\boldsymbol{r}' = -\frac{1}{\pi} (\boldsymbol{K}^* - \boldsymbol{K}^{R*}) \boldsymbol{t}$\;
  $\tau = \sum \limits_{i \in I_{\text{con}}} \widetilde{V}_i^* t_i/\sum \limits_{i \in I_{\text{con}}} r'_i t_i$ \tcp*{Calculate the step size}

  $\boldsymbol{J}^* \leftarrow \boldsymbol{J}^* - \tau {\bf t}$ \tcp*{Correct $\boldsymbol{J}^*$ along the conjugate direction}

  {\bf Set} all negative $J_i^*$ to zero\;

  {\bf Update} $I_{\text{ol}} = \{i|J_i^* = 0, \widetilde{V}_i^* < 0, i = 1, \cdots, n \}$\;

  $J_{i \in I_{\text{ol}}}^* \leftarrow J_{i \in I_{\text{ol}}}^* - \tau \widetilde{V}_{i \in I_{\text{ol}}}^*$\tcp*{Correct negative $\widetilde{V}_i^*$}
    
  $\epsilon_{\text{p}} = \sqrt{\sum \limits_{i = 1}^n (J_i^* - J_i^{^*\text {old}})^2}/\sqrt{\sum \limits_{i = 1}^n J_i^{*2}}$ \tcp*{Error}
  
  $\boldsymbol{J}^{*\text{old}} = \boldsymbol{J}^*$, $I_{\text{con}} = \{ i |J_i^* > 0, i = 1, \cdots, n \}$\;
 }
$\boldsymbol{V}^* = \boldsymbol{h}^* - \widetilde{\boldsymbol{V}}^* $\;
$I^* = \Delta_x^* \sum \limits_{i = 1}^n J_i^*$
\end{algorithm}

\end{document}